\documentclass[twocolumn,showpacs,preprintnumbers,amsmath,amssymb]{revtex4}

\usepackage{graphicx}
\usepackage{dcolumn}
\usepackage{bm}

\newcommand\ignore[1]{}
\def\one{{\,\hbox{1\kern-.8mm l}}}

\newcommand{\Cset}{{\,\,{{{^{_{\pmb{\mid}}}}\kern-.45em{\mathrm C}}}}}

\newcommand{\be}{\begin{equation}}
\newcommand{\ee}{\end{equation}}
\newcommand{\bea}{\begin{eqnarray}}
\newcommand{\eea}{\end{eqnarray}}

\providecommand{\lsim}{\lesssim}

\def\a{\alpha}

\begin{document}

\title{Solution of the cosmological constant problem within holographic cosmology}

\author{Horatiu Nastase$^{1}$}\email{horatiu.nastase@unesp.br}
\affiliation{${}^{1}$Instituto de F\'{i}sica Te\'{o}rica, UNESP-Universidade Estadual Paulista, Rua Dr. Bento T. Ferraz 271, Bl. II, 
S\~ao Paulo 01140-070, SP, Brazil}

\date{\today}

\begin{abstract}
Within the holographic cosmology paradigm, specifically the model of McFadden and Skenderis, but more generally than that, 
the cosmological constant is found to naturally flow from a large value, to a small value, which can even be as low as the needed
 $\sim 10^{-120}$ times the original, along with the 
dual RG flow. Within this context then, the cosmological constant problem is mapped to a simple quantum field theory property, even though the exact mechanism 
for it in gravity is still obscure. I consider several examples of gravity duals to explain the mechanism.

\end{abstract}


\maketitle

\section{Introduction}

In cosmology, the $\Lambda$CDM model with inflation has become a sort of standard model of cosmology, and experimentally, the measured acceleration of the Universe
implies that $\Lambda$ is at least approximately constant in time, more like a cosmological constant than a quintessence. Inflation also implies that at the beginning of the 
Universe, the potential energy was approximately constant for a period of time, within the effective field theory approach. 
In this paradigm however, we have the cosmological constant problem: the effective field theory approach means that one should be able to compute a vacuum energy 
perturbatively and, even in a supersymmetric model with low energy breaking of supersymmetry, this would generate an energy density of the order of the cut-off scale
(the susy breaking scale in the supersymmetric scenario), which would gravitate, and drastically over-close the Universe, being many orders of magnitude above the 
observed value today. This would suggest that perhaps there is something wrong with the intersection of perturbative quantum field theory and gravity, which doesn't 
allow us to solve the problem. 

But over the last 10 years or so, there is a new and larger paradigm, including inflation, that could in principle address this problem: that of holographic cosmology. 
Holographic cosmology as a general strategy was defined in \cite{Maldacena:2002vr}, based on the proposed relation between the wave function of the Universe
in cosmology with a given boundary value $\phi$ for the fields, $\Psi[\phi]$, and the partition function on the boundary with source $\phi$, $Z[\phi]$, as $\Psi[\phi]
=Z[\phi]$. A concrete proposal for such a holographic correspondence between cosmology and a field theory was only developed in \cite{McFadden:2009fg}
(there were other attempted constructions, like for instance the top-down one in \cite{Brandenberger:2016egn,Ferreira:2016gfg}, based on the earlier 
\cite{Awad:2008jf}, but they are not so well developed, or general). In the model, a holographic computation, based on the methods of 
\cite{Skenderis:2000in,Papadimitriou:2004ap,Papadimitriou:2004rz}, relates the correlators of fluctuations of $h_{ij}$ observed in the sky, $\langle \delta h_{ij}
\delta h_{kl}\rangle$, with boundary correlators $\langle T_{ij} T_{kl}\rangle$, in a super-renormalizable 2+1 dimensional field theory with "generalized 
conformal symmetry". The holographic calculation is equivalent with a derivation from Maldacena's $\Psi[\phi]=Z[\phi]$ ansatz above 
\cite{Afshordi:2017ihr}. Since the ansatz was originally meant for inflation, one can see that the holographic cosmology paradigm of \cite{McFadden:2009fg} includes 
inflation. The specific new model considered from this paradigm is a phenomenological approach, writing the most general 2+1 dimensional field theory
of $SU(N)$ gauge fields, scalars and fermions consistent with the symmetries, and considered in perturbation theory, so that the dual cosmological theory would be
non-perturbative, and not addressable by standard methods.  This model is thus continuously connected with inflation, but different from it, and with different predictions. 
In two remarkable papers \cite{Afshordi:2016dvb,Afshordi:2017ihr}, it was further shown that the new model, although predicting different functional forms for the 
power spectra of the CMBR, is experimentally as good as $\Lambda$CDM with inflation: in a Bayesian approach, fitting the data against the predictions of the 
two models gives results with one sigma of each other: the difference in $\chi^2$'s, which are of the order of 850, is of about 0.5. 
Moreover, in work to be published soon \cite{HoratiuKostas1,HoratiuKostas2}, it is found that the holographic cosmology paradigm also solves the usual 
pre-inflationary problems with cosmology as well as inflation.

It is then a natural question to ask: can one solve, or at least map holographically to a solved problem, the cosmological constant problem within holographic cosmology? 
In this paper, I will give an answer in the affirmative, and show that there is a natural flow in the potential energy, going along with the RG flow, between a large 
energy, corresponding to the analog of inflation, until a small energy, corresponding to the dark energy today, and that the huge relative factor, of about $10^{-120}$, 
can be naturally obtained in field theory.

The paper is organized as follows. In section 2 I present the main argument of the paper, that the cosmological constant is defined by the RG flow in 2+1 dimensions, 
based on the general principles of holographic cosmology. In section 3 I give examples of holographic duals, building up to the result for the relevant cosmology, 
and show that they are consistent with the proposed principle. In section 4 I conclude.

\section{Field theory dual to the cosmological constant, and its flowing}

Within the holographic cosmology paradigm, a cosmology with a scale parameter dependence $a(t)$ and a scalar field $\phi(t)$ is Wick rotated
to a "domain wall" background,
\be
ds^2=dr^2+a(r)^2d\vec{x}^2\;,\;\;\; \phi=\phi(r)\;,
\ee
i.e., we have a "domain wall/cosmology correspondence", which is considered to be the gravity dual to a phenomenological Euclidean 3 dimensional field 
theory via usual holography. The field theory contains $SU(N)$ adjoint fields, scalars $\phi^I$ and fermions $\psi^K$, and has the property of "generalized conformal structure", 
meaning that we can rescale fields and couplings such that the only dimensionful coupling is the overall gauge coupling factor $1/g^2$. Then one can show that quantum 
corrections organize themselves in terms of the dimensionless effective coupling 
\be
g^2_{\rm eff}=\frac{g^2N}{q}\;,
\ee
where $q$ is the momentum scale. One can calculate field theory quantities in perturbation theory, like $T_{ij}$ correlators, which will give log-corrected polynomials 
in $g^2_{\rm eff}$, as usual, and then relate these quantities holographically to cosmological quantities (like $h_{ij}$ correlators in the case of $T_{ij}$).

The RG flow of the 3 dimensional field theory is thus dominated by the dependence on $q$ of $g^2_{\rm eff}$. In \cite{HoratiuKostas1,HoratiuKostas2} it was found that 
the same RG flow gives a dilution effect, responsible for solving the monopole, flatness, and smoothness and horizon problems of pre-inflationary cosmology. 
It is then natural to ask whether the same effect cannot also solve the cosmological constant problem?

In order to answer that, we would need to find the field theory quantity that corresponds to the quantum cosmological constant, or more precisely the (gravitating) 
one-loop quantum corrections for matter. Normally, these would be $\a'$ string corrections to the gravitational theory. 

Considering the best understood example as a first try, of ${\cal N}=4$ SYM vs. string theory in $AdS_5\times S^5$, we have
\be
\frac{1}{\lambda}=\frac{1}{g^2_{YM}N}=\frac{\a'^2}{R^4}\sim \a'^2{\cal R}^2\Rightarrow \a' {\cal R}\sim 
\frac{1}{\sqrt{\lambda}}\;,
\ee
where $R$ is the radius of $S^5$ and of $AdS_5$, and ${\cal R}$ is the Ricci scalar.

In holographic cosmology, gravity has an (approximate) cosmological constant (at least in the extreme UV, for the beginning of inflation, and in the extreme IR, for the far future) 
and/or scalar potential, so we consider
\be
S_{\rm gravity}=\frac{M^2_{\rm Pl}}{2}\int d^4x (R+2\Lambda)\;,
\ee
to which we add the matter action. Then the Einstein equations 
\be
{\cal R}_{\mu\nu}-\frac{1}{2}g_{\mu\nu}{\cal R}=\Lambda g_{\mu\nu}+8\pi G_N T_{\mu\nu}
\ee
imply for the trace in $d$ spacetime dimensions
\be
\frac{2-d}{2}{\cal R}=d\Lambda +8\pi G_N T.
\ee

As a consequence, in this simple case, we can write 
\be
\frac{\Lambda}{M^2_{\rm Pl}}\lsim \frac{\cal R}{M_{\rm Pl}^2}\sim \frac{1}{\sqrt{\lambda}}.
\ee

But in the case of usual holographic cosmology things are more complicated. First of all, in terms of the background solution, the large time region, when the Universe is 
large, corresponds to the UV of the field theory, $q\rightarrow\infty$, in which case the field theory is perturbative. Since holographic cosmology is considered mostly as a 
replacement for inflation, mostly tested through its imprint on the CMBR, we note that the nonperturbative regime for the CMBR was found in \cite{Afshordi:2016dvb,Afshordi:2017ihr}
to correspond to modes $l\lsim 30$, which were the earliest CMBR modes to be created during inflation. The modes created later are perturbative. 

On the other hand, the large time region has small ${\cal R}$ and $\Lambda$, and the $q\rightarrow \infty$ region corresponds to $g^2_{\rm eff}\rightarrow 0$, so 
we can only have ${\cal R}/M^2_{\rm Pl}\sim (g_{\rm eff})^p$, with $p>0$, unlike the $p=-1/2$ obtained for ${\cal N}=4$ SYM. We will in fact check that this is the case
in the next section. 

Assuming this behaviour, we obtain 
\be
\frac{\Lambda}{M^2_{\rm Pl}}\lsim \left(\frac{g^2N}{q}\right)^p\;,
\ee
and this means that {\em we obtain the natural flowing of the cosmological constant from a high one during inflation to a low one today simply as a consequence of 
the dimensional RG flow in field theory}. And the large suppression factor of $10^{-120}$ is simply due to a similarly large flowing in scales $q$, necessary to 
come from inflationary times to now. In this case, we have assumed that $q\propto t^r$, with $r>0$, so that UV in field
theory corresponds to late times in cosmology.

We can state this fact as a principle, the {\bf principle} that the quantum cosmological constant, for a 3+1 dimensional quantum gravity theory in a FLRW cosmology, 
is holographically related to the evolution of the dual 2+1 dimensional scale in quantum field theory: low $\Lambda$ corresponds to 
high $q$, which in turn means low $g^2_{\rm eff}$. 

In the following section we try to check if this is true through examples. 

\section{Examples of gravity duals}

{\bf Holographic dual of D$\tilde p$-branes}

The simplest example, and one on which the construction of \cite{McFadden:2009fg} is based, is of the holographic dual derived from (non-conformal, so $\tilde p\neq 3$) 
D$\tilde p$-branes in 10 dimensions, defined in  \cite{Itzhaki:1998dd}. In this case, the effective coupling is $g^2_{\rm eff}=g^2_{YM}N U^{\tilde p-3}$, where
$U=r/\a'$ corresponds to the momentum scale, $q$ of the previous section, 
so perturbation theory, $g^2_{\rm eff}\ll 1$, means that $U\gg (g^2_{YM}N)^{1/(3-\tilde p)}$ 
for $\tilde p<3$ and 
$U\ll 1/(g^2_{YM}N)^{1/(\tilde p-3)}$ for $\tilde p>3$.

The extremal D$\tilde p$-brane solution is 
\bea
ds^2&=& f_{\tilde p}^{-1/2}dx_{||}^2+f_{\tilde p}^{1/2}dx_{\perp}^2\cr
e^{-2(\phi-\phi_\infty)}&=&f_{\tilde p}^{\frac{\tilde p-3}{2}}\cr
\a'^2f_{\tilde p}&=&\a'^2+\frac{d_{\tilde p} g^2_{YM}N}{U^{7-\tilde p}}\;,
\eea
which leads, for $\tilde p=2$, to 
\bea
e^{\phi}&\sim & \left(\frac{g^2_{YM}N}{U}\right)^{5/4}\frac{1}{N}\cr
\frac{ds^2}{\a'}&\simeq &U^2\sqrt{\frac{U^{3-\tilde p}}{g^2_{YM}N d_{\tilde p}}}dx_{||}^2
+\sqrt{\frac{g^2_{YM}Nd_{\tilde p}}{U^{3-\tilde p}}}
\frac{dU^2}{U^2}\cr
&&+\sqrt{\frac{g^2_{YM}N d_{\tilde p}}{U^{3-\tilde p}}}d\Omega_{8-\tilde p}^2\cr
&=& \frac{U^2}{R^2}dx_{||}^2+R^2\frac{dU^2}{U^2}+R^2d\Omega_{8-\tilde p}^2\Rightarrow \cr
\a' {\cal R}&\sim & \frac{1}{\sqrt{\lambda_{\rm eff}}}\sim 
\sqrt{\frac{U^{3-\tilde p}}{g^2_{YM}N}}\rightarrow \sqrt{\frac{U}
{g^2_{YM}N}}.
\eea

The validity of supergravity conditions in the $\tilde p=2$ (D2-brane) case, relevant for the holographic cosmology construction (the field theory is 2+1 dimensional)
 are $e^\phi\ll 1$ and $\a' {\cal R}\ll 1$, which give $\lambda_{\rm eff}\ll 1$ and 
$N\gg 1$, or more precisely
\bea
&& 1\ll g^2_{\rm eff}\ll N^{\frac{4}{7-\tilde p}}\rightarrow N^{4/5}\Rightarrow\cr
&& g^2_{YM}N^{1/5}\ll U \ll g^2_{YM}N.
\eea
For $g^2_{YM}N\ll U$ (at the largest $U$), the gravity solution is not valid, and perturbative SYM (with $\frac{g^2_{YM}N}
{U}\ll 1$) is to be used instead. 

Note however that in this case, if we were to redefine coordinates of this {\em string frame} solution 
to put it in the FLRW cosmology form (which is of course not what one is instructed to do, the Einstein frame cosmology 
is what really matters), 
\be
ds^2=U^{\frac{7-\tilde p}{2}}dx^2_{||}+\frac{dU^2}{U^{\frac{7-\tilde p}{2}}}=dt^2+a^2(t)dx_{||}^2\;,
\ee
we would obtain 
\be
t\sim U^{\frac{\tilde p-3}{4}}\;,\;\;\;
a(t)\sim t^{\frac{7-\tilde p}{\tilde p-3}}.
\ee

Specifically, for the relevant case of $\tilde p=2$, we would  obtain 
\be
t\sim U^{-1/4}\;,\;\;\;
a(t)\sim t^{-5}\;,
\ee
so large time $t$ means small $U$, but it also means small $a(t)$. That means that  there is no contradiction with the holographic 
cosmology analysis from the previous section, just that now, from the point of view of this {\em string frame} 
FLRW metric, we have a {\em contracting} phase. 
Then large time corresponds to the IR of the field theory (apparently the opposite of our analysis), but only because it
is also the place where the Universe shrinks to zero size. Then the correct statement is that the IR of the field theory corresponds to the small (spatial) Universe 
region, and is the same as in our analysis. In this string frame, we would obtain 
\be
\frac{\Lambda_S}{M_{\rm Pl}^2}\lsim \frac{{\cal R}_S}{M_{\rm Pl}^2}\propto U^{1/2}\propto \frac{1}{t^2}
\propto a(t)^{2/5}\;,
\ee
which contradicts the behaviour with $U$ ($=q$) proposed, but not with $t$. We can therefore safely consider it to be an 
artifact of the strange contracting Universe.

But moreover, what we should have been analyzing is the {\em Einstein frame} FLRW cosmology. In Einstein frame, 
\bea
&&{\cal R}_E\sim e^{\phi/2}{\cal R}_S+...\sim \frac{\lambda_{\rm eff}^{1/8}}
{\sqrt{N}}=\frac{1}{\sqrt{N}}\left(\frac{g^2N}{U}
\right)^{1/8}\Rightarrow\cr
&&\frac{\Lambda_E}{M_{\rm Pl}^2}\lsim \frac{{\cal R}_E}{M_{\rm Pl}^2}\propto
\left(\frac{g^2N}{U}
\right)^{1/8}\;,
\eea
in other words, $p=1/8$. And then, in this case, in Einstein frame and after KK reduction on the sphere,  
one obtains correctly expanding FLRW cosmology with 
$a(t)\propto t^7$ \cite{Kanitscheider:2008kd} (and $t$ increasing exponentially with $U$). 
Of course, as we saw, the {\em late time} (large $t$, i.e., large $U$) cosmology in this case is described by perturbative 
SYM, which means that in this case, in order to describe the real world, the theory would have to somehow transition further 
into a gravitational description, like the one we have in the Universe (at the very least) after Big Bang Nucleosynthesis, 
or after reheating, in the inflationary paradigm. But we can put this unusual feature
down to the strange behaviour in string frame.
There is still the issue of what does it mean for the field theory to have a drop of $\sim 10^{-120}$ in 
$\Lambda$ from the Planck scale to today. Assuming a radiation dominated (R.D.) Universe, as it is true for 
most of its evolution, since $\Lambda\propto H^2\propto a(t)^{-4}\propto T^4$, that amounts to a drop
in temperature 
$T$ from $T_{\rm Pl}\sim 10^{19}GeV$ to about $10 meV$ (a factor of $10^{-30}$). 
Since we don't have a model for the 
transition to the standard R.D. Universe, we can only argue about the holographic 
cosmology period, which for the 
D2-brane model gives $U\propto a(t)^s\propto T^{-s}$, with $s=4/5$. In the R.D. Universe, we expect 
a different value for $s$, but still $\sim {\cal O}(1)$. 
Then the drop in $T$ by $10^{-30}$ means an increase in energy $U$, along the inverse
RG flow, by a factor of $10^{24}$. 
Since the dual field theory has no cut-off (there is no Planck scale for gravity
to cut off the increase in energy), this is not unreasonable. 

{\bf Compactified NS5-branes}

The next, more relevant, case is the holographic dual of a ${\cal N}=1^*$ theory, for which the minimal supersymmetry implies a running coupling constant and
RG flow already in 3+1 dimensions. That case (NS5-branes compactified on $S^2$ with a twist)
was studied in \cite{Maldacena:2000yy}, however, I will not review it here, and move directly on to the similar 
2+1 case (NS5-branes compactified on $S^3$ with a twist), relevant for our purposes, defined in \cite{Maldacena:2001pb}. The gravity dual solution is not defined 
analytically throughout all values for the radial coordinate, but only at the extremes (UV and IR).

In this case, {\em in the UV of the solution} (corresponding to late time cosmology), where $\rho\rightarrow \infty$, 
we  have
\be
g_{s,eff}=e^\phi\sim e^{-\rho}\rightarrow 0\;,
\ee
meaning that  the solution has no string corrections. But also, in Einstein frame, 
\be
\a' {\cal R}_E\sim \sqrt{g_s}\ll 1\;,
\ee
and moreover also in string frame,  we have
\be
\a' {\cal R}_S\sim \frac{1}{N R^2(\rho)}\sim \frac{1}{N\rho}\rightarrow 0.
\ee

Since we have 
\be
ds^2_E\sim e^{-\phi/2}dx_{2+1}^2+...\Rightarrow ds^2_S\sim dx_{2+1}^2+...\;,
\ee
we obtain
\be
\a' {\cal R}_S \ll 1\;,\;\;\;
\frac{{\cal R}_E}{M^2_{\rm Pl}}\sim \frac{e^{\phi/2}}{M^2_{\rm Pl}}\ll 1\;,
\ee
because $e^\phi\ll 1$ ($g_s\ll 1$). 

Thus we have simultaneously, ($g_s\ll 1, \a'{\cal R}_S\ll 1$), or ${\cal R}_E/M^2_{\rm Pl}\ll 1$. Either way, string corrections to the geometry are small in the 
UV of the solution, corresponding to the late time cosmology. More importantly, according to our analysis, 
since $g_{s,eff}\rightarrow 0$ as well as $\frac{\cal R}{M_{\rm Pl}^2}\rightarrow 0$, we must also have, in both frames, 
\be
\frac{{\cal R}}{M_{\rm Pl}^2}\sim (g^2_{\rm eff})^p\Rightarrow \frac{\Lambda}{M^2_{\rm Pl}}\lsim \left(\frac{g^2N}{q}\right)^p\;,\;\; p>0\;,
\ee
as advertised. We see that eliminating the unphysical ``contracting Universe'' in string frame from 
the previous example leads to a behaviour fully consistent with what we wanted. However, in this case, the UV of the field 
theory is not quite the free SYM, since we have also the KK modes on the $S^3$ on which the NS5-branes are wrapped, 
so the analysis is more complicated. But that was to be expected, since there is no known gravity dual to pure ${\cal N}=1$
SYM.

{\bf Generic holographic cosmology}

We finally consider the generic case of a FLRW cosmology gravity dual, with metric
\be
ds^2=-dt^2+a^2(t)d\vec{x}^2.
\ee
It is to be understood in the sense of the phenomenological approach of \cite{McFadden:2009fg}, though a 
specific ``top-down'' implementation, with a correct continuation into the current gravitational cosmology, could also be
imagined in principle.

Then the Ricci scalar is 
\be
{\cal R}=6\frac{\ddot a}{a}+6\frac{\dot a^2}{a^2}.
\ee
In the AdS case, when $a(t)\sim e^{kt}$, we get $R=6k^2$, independent of $t$. 

But for a power law cosmology, $a(t)\sim t^n$, as we are most interested in, we obtain the Ricci scalar
\bea
{\cal R}&=&\frac{n(2n-1)}{t^2}\propto \frac{1}{[a(t)]^{2/n}}\propto \frac{1}{q^{2r}}\Rightarrow\cr
\frac{\Lambda}{M_{\rm Pl}^2}&\lsim &\frac{C}{q^p}\;,
\eea
if the field theory scale $q$ is related to holographic time by $t\sim q^r$, with $r>0$, where $C$ is a constant. 
But as we saw from the previous
example of the ${\cal N}=1^*$ SYM theory in 2+1 dimensions, and to the Einstein frame for D2-branes,
the UV of the theory (large $q$) corresponds to large times $t$ for an expanding cosmology, 
so indeed $p=2r>0$.

Note that in the D$\tilde p$-brane case in string frame we still get $\a' {\cal R}\propto 1/t^2$, 
just that now this is also 
$\propto U^{\frac{3-\tilde p}{2}}$ and $\propto 1/\sqrt{g^2_{YM}N}$, as was found by  \cite{Itzhaki:1998dd}, and we can now check.

\section{Conclusions}

In this paper, I have shown that within the general holographic cosmology framework, where a 4 dimensional cosmology has the time direction reinterpreted, via a Wick 
rotation (domain wall/cosmology correspondence) as a radial holographic direction, mapping to a 2+1 dimensional field theory, the evolution of the approximate cosmological 
constant from a very large one during inflation, or its nongeometric analog, to a very small one (dark energy) now, is dual to the evolution with the RG scale $q$ of the
field theory. The relation one obtains is of the type 
$\frac{\Lambda}{M^2_{\rm Pl}}\lsim \left(\frac{g^2N}{q}\right)^p$, with $p>0$.

I have shown that in the D$\tilde p$-brane holographic dual (specifically for $\tilde p=2$) 
case, the Einstein frame behaviour is correct, and while the string frame one 
seems to go the opposite way, it is only because the related "cosmology" is contracting. 
The more relevant case of the ${\cal N}=1^*$ dual in 2+1 dimensions in \cite{Maldacena:2001pb} is consistent with the analysis, and a general holographic cosmology shares 
the features proposed.

Of course, I have not solved the cosmological constant problem {\em in gravity}, I have only mapped it via holography to an understood problem in field theory.

\section*{Acknowledgements}

I thank Kostas Skenderis for many useful discussions on the subject of  holographic cosmology, and in particular of RG flows within it.
My work is supported in part by CNPq grant 304006/2016-5 and FAPESP grant 2014/18634-9. I would also like to 
thank the ICTP-SAIFR for their support through FAPESP grant 2016/01343-7, and the University of Cape Town for hospitality when this work was being 
finalized.

\end{document}